\documentclass[%
a4paper,prl,final,superscriptaddress,twocolumn
]{revtex4-1}

\usepackage{graphicx}
\usepackage{dcolumn}
\usepackage{bm}
\usepackage{siunitx}
\usepackage{tabularx}
\usepackage{latexsym}
\usepackage{esvect}
\usepackage[colorlinks = true,
            linkcolor = blue,
            urlcolor  = blue,
            citecolor = blue,
            anchorcolor = blue]{hyperref}

\usepackage{mathtools}
\usepackage{amssymb}
\usepackage{bm}
\usepackage{color}
\usepackage{braket}
\usepackage{dsfont}
\usepackage{wasysym}

\begin{document}

\title{Magnetic field independent sub-gap states in hybrid Rashba nanowires}

\author{Christian J\"unger}
\thanks{These two authors contributed equally.}
\affiliation{Department of Physics, University of Basel, Klingelbergstrasse 82, CH-4056 Basel, Switzerland}

\author{Rapha\"elle Delagrange}
\thanks{These two authors contributed equally.}
\affiliation{Department of Physics, University of Basel, Klingelbergstrasse 82, CH-4056 Basel, Switzerland}

\author{Denis Chevallier}
\affiliation{Department of Physics, University of Basel, Klingelbergstrasse 82, CH-4056 Basel, Switzerland}

\author{Sebastian Lehmann}
\affiliation{
 Division of Solid State Physics and NanoLund, Lund University, S-221 00 Lund, Sweden
}

\author{ Kimberly A. Dick}
\affiliation{
 Division of Solid State Physics and NanoLund, Lund University, S-221 00 Lund, Sweden
}
\affiliation{Center for Analysis and Synthesis Lund University, S-221 00 Lund, Sweden.}

\author{Claes Thelander}
\affiliation{
 Division of Solid State Physics and NanoLund, Lund University, S-221 00 Lund, Sweden
}

\author{Jelena Klinovaja}
\affiliation{Department of Physics, University of Basel, Klingelbergstrasse 82, CH-4056 Basel, Switzerland}

\author{Daniel Loss}
\affiliation{Department of Physics, University of Basel, Klingelbergstrasse 82, CH-4056 Basel, Switzerland}

\author{Andreas Baumgartner}
\thanks{Email: andreas.baumgartner@unibas.ch}
\affiliation{Department of Physics, University of Basel, Klingelbergstrasse 82, CH-4056 Basel, Switzerland}
\affiliation{Swiss Nanoscience Institute, University of Basel, Klingelbergstrasse 82, CH-4056, Basel, Switzerland}

\author{Christian Sch\"onenberger}
\affiliation{Department of Physics, University of Basel, Klingelbergstrasse 82, CH-4056 Basel, Switzerland}
\affiliation{Swiss Nanoscience Institute, University of Basel, Klingelbergstrasse 82, CH-4056, Basel, Switzerland}

\date{\today}
\begin{abstract}
Sub-gap states in semiconducting-superconducting nanowire hybrid devices are controversially discussed as potential topologically non-trivial quantum states. One source of ambiguity is the lack of an energetically and spatially well defined tunnel spectrometer. Here, we use quantum dots directly integrated into the nanowire during the growth process to perform tunnel spectroscopy of discrete sub-gap states in a long nanowire segment. In addition to sub-gap states with a standard magnetic field dependence, we find topologically trivial sub-gap states that are independent of the external magnetic field, i.e. that are pinned to a constant energy as a function of field. We explain this effect qualitatively and quantitatively by taking into account the strong spin-orbit interaction in the nanowire, which can lead to a decoupling of Andreev bound states from the field due to a spatial spin texture of the confined eigenstates.
\end{abstract}
\pacs{Valid PACS appear here}
\maketitle
Semiconducting Nanowires (NWs) often have a strong inherent spin-orbit interaction (SOI), which for example, lies at the core of fast electrical manipulation of spin-orbit qubits \cite{Nadj-Perge2010,Berg2013}. Combining such NWs with an \textit{s}-wave superconductor (SC) can in addition give rise to topologically non-trivial superconductivity and to topologically protected bound states at the ends of the proximitized NW region, for example Majorana bound states (MBSs) \cite{Mourik2012,Albrecht2016,Zhang2018}, potentially useful for topological quantum computation \cite{Nayak2008,Alicea2011}.
More insight into MBSs can be gained by deterministic tunnel spectroscopy, providing information on the lifetimes \cite{Leijnse2011}, parity \cite{Gharavi2016}, and spin-texture \cite{Chevallier2018,Prada2017} of MBSs. Controlled tunnel spectroscopy using in-situ grown quantum dots (QDs) was for example used to probe the superconducting proximity effect in NWs \cite{Juenger2019}.
%
%
%
\begin{figure}[b!]
\includegraphics[width=\columnwidth]{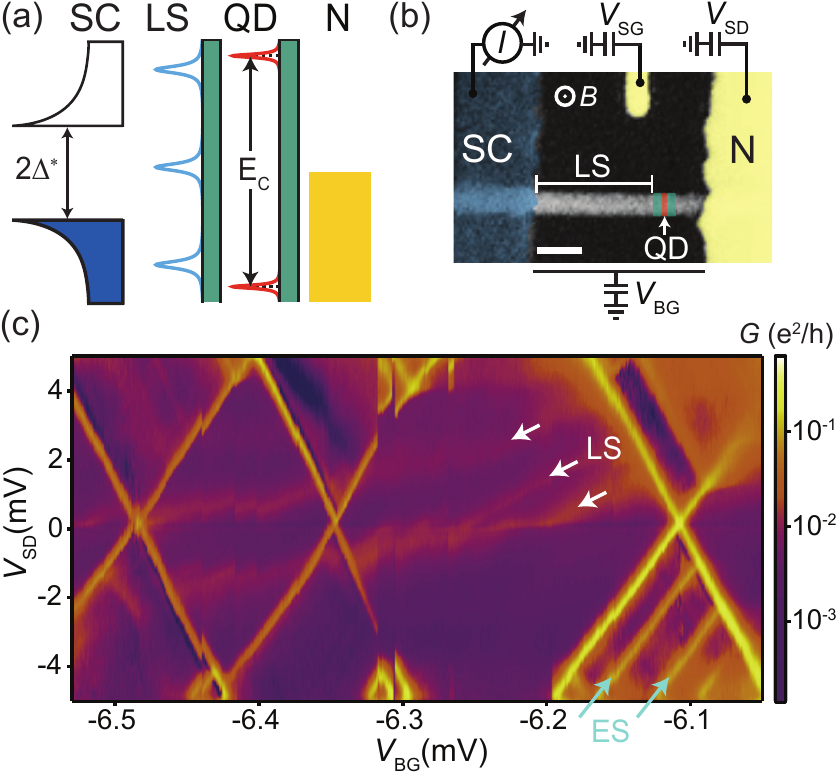}
\caption{\label{fig:setup}(a) Energy diagram of the considered system, with a quantum dot (QD, red states), a lead segment (LS) with discrete energy levels (blue) strongly coupled to a superconducting (SC) and a normal metal electrode (N). (b) False color scanning electron micrograph of a representative device, made of an InAs nanowire with the integrated QD (red) formed by two tunnel barriers (green) grown in the wurtzite phase, located $\ell_{LS}$~$\approx$~\SI{220}{nm} from SC (scale bar: \SI{100}{nm}). The measurement setup is shown schematically on top. The external magnetic field $B$ is applied out-of-plane, perpendicular to the NW axis. (c) Differential conductance, $G$, as a function of the backgate voltage, $V_{\text{BG}}$, and the voltage bias, $V_{\text{SD}}$, in the normal state ($B$~=~\SI{50}{\milli \tesla}). LS resonances inside the Coulomb blockade regime of the QD are pointed out by white arrows, QD excited states (ES) by green arrows.}
\end{figure}
Here, we use the same material platform to perform transport spectroscopy on a segment of a Rashba NW carrying discrete superconducting sub-gap states, the nature of which depends strongly on the SOI in the NW.
The energy of such states, typically follows the energy of the superconductor which closes with increasing magnetic field \cite{Lee2012, Chang2013}. 
In contrast, we demonstrate here that a strong Rashba SOI in the NW can result in field independent, topologically trivial Andreev bound states (ABSs). We explain these findings by a decoupling of a finite length ABS from the external magnetic field $B$ due to a spatial spin texture and the vanishing of the corresponding effective magnetic moment \cite{Reeg2018,Trif2008}.
We use these results to estimate the Rashba spin-orbit strength in the relatively long NW segment, in contrast to previous experiments based on studying the singlet - triplet splitting in a QD \cite{Fasth2007}, or on weak anti-localization measurements in the diffusive limit \cite{Scheruebl2016}.\\
\begin{figure*}[t]
\includegraphics[width=2\columnwidth]{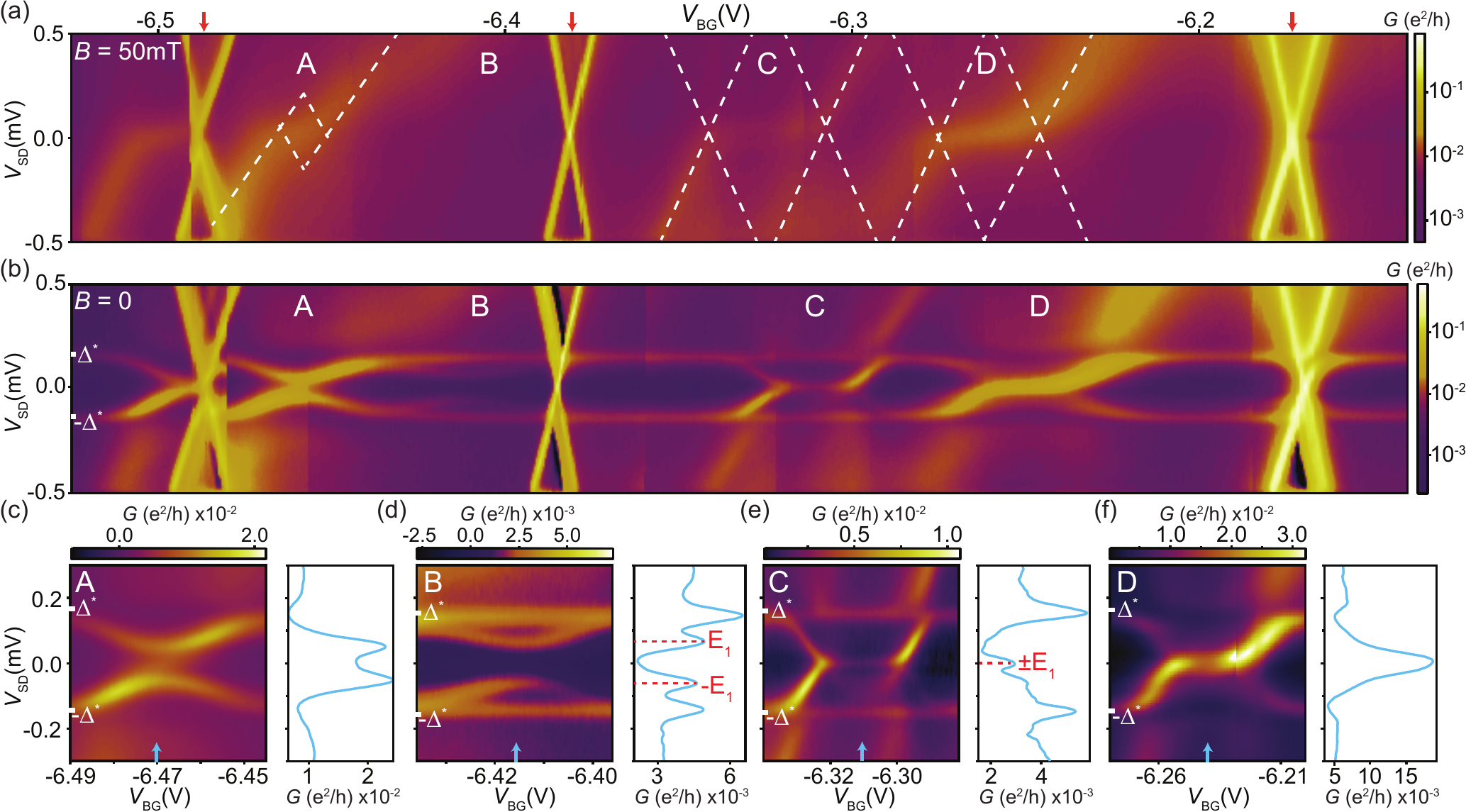}
\caption{\label{fig:SpectroABS} (a) $G$ as a function of $V_{\text{BG}}$ and $V_{\text{SD}}$ in the normal state ($B$ = \SI{50}{\milli \tesla}) and (b) in the superconducting state  ($B$ = 0). (c-f) Detailed measurements of the regions labeled in (a) and (b), including cross sections taken at the constant $V_{\text{BG}}$ indicated by a blue arrow.}
\end{figure*}
%
To perform tunnel spectroscopy we use deterministically grown wurtzite (WZ) segments in zincblende (ZB) InAs NWs \cite{Dick2010,Lehmann2013}. The two WZ segments act as hard wall barriers for electrons, as the ZB and WZ band structure align with a conduction band offset of $\sim$\SI{100}{\milli eV} \cite{Nilsson2016,Chen2017,Juenger2019}, which results in the formation of a QD in the ZB segment between the barriers. This QD allows us to probe the tunneling density of states (DOS) in the bare NW ``lead segment'' (LS) by quantum tunneling.
As most of the bias develops across the integrated QD, the measured differential conductance, $G$, directly maps the excitation spectrum in the LS, see Fig.~\ref{fig:setup}~(a).\par
A false color scanning electron micrograph of a representative device is shown in Fig.~\ref{fig:setup}~(b). The QD (red, \SI{20}{nm}) is defined by two WZ segments (green, \SI{30}{nm}) within the ZB InAs NW. The normal metal electrode (yellow) consists of titanium/gold (Ti/Au, \SI{5}{nm}/\SI{70}{nm}), whereas the superconducting electrode (blue) is made of titanium/aluminum (Ti/Al, \SI{5}{nm}/\SI{80}{nm}), both defined by standard electron beam lithography, with a distance of $\sim$ \SI{350}{nm} between both contacts. In the device discussed below, the QD is located about $\ell_{LS}$~$\approx$~\SI{220}{nm} from the Al electrode, leaving a bare NW segment of this length between the QD and SC. Specifically, the QD is not directly coupled to the SC \cite{Juenger2019}. The backgate voltage, $V_{\text{BG}}$, tunes the chemical potential in both the QD and the LS.\par
Figure~\ref{fig:setup}~(c) shows an overview measurement of the differential conductance $G$, as a function of $V_{\text{BG}}$ and the bias voltage $V_{\text{SD}}$, with regular, very sharp Coulomb blockade (CB) diamonds of the QD in the normal state $B =$~\SI{50}{\milli \tesla} (charging energy $E_{\text{C}}$~$\approx$~\SI{5}{\milli eV}, level spacing $\varepsilon$~$\approx$~\SIrange{1}{2}{\milli eV}). In the CB regime, the charge state of the QD is fixed and the dot can be thought of as a single tunnel barrier, with the electronic transport mediated by cotunneling \cite{Franceschi2001}.
Here, we observe broad discrete resonances pointed out by white arrows in Fig.~\ref{fig:setup}~(c), running through the CB diamonds of the integrated QD. These resonances are not cotunneling lines mediated by excited states of the QD (ES, green arrows in Fig.~\ref{fig:setup}~(c)), since they display a different lever arm and can be tuned independently by the sidegate voltage, $V_{SG}$, as discussed in the Supplemental Material (see Fig.~S1). We therefore interpret these states as bound-states that form in the LS of the NW and refer to them as ``LS resonances''.\par
%
%
%
%
%
%
As shown in Fig.~\ref{fig:SpectroABS}~(a) for small bias voltages and in the normal state, we detect a set of resonances, following a pattern resembling a broad spin-1/2 Kondo resonance \cite{Cronenwett1998,Goldhaber-Gordon1998}.
In particular, we find several conductance ridges centered at zero bias, each over a gate-voltage range of  $\sim$ \SI{25}{\milli eV}, best seen in the regions labeled A and D, and weaker in C. The associated approximate CB diamonds are shown as dashed lines, which suggest an addition energy $E_{\rm{add}} \approx$~\SIrange{0.2}{0.5}{\milli eV}, i.e. $\sim$~10 times smaller than the addition energy of the integrated QD, consistent with the longer LS.\par
In the superconducting state ($B$ = 0), we find an induced gap of $\Delta^*$ $\approx$ \SI{150}{\micro eV}, very similar to previous experiments using evaporated Al \cite{Juenger2019}. The LS resonances of the normal state develop into sharp discrete sub-gap states, as seen in  Fig.~\ref{fig:SpectroABS}~(b), similar to those observed in other experiments on hybrid QD systems \cite{Eichler2007,Pillet2010,Lee2012,Chang2013,Schindele2014,Jellinggaard2016,Kim2013,Gramich2017,Grove-Rasmussen2018}.\par
%
%
%
%
%
%
\begin{figure}[t]
\includegraphics[width=1\columnwidth]{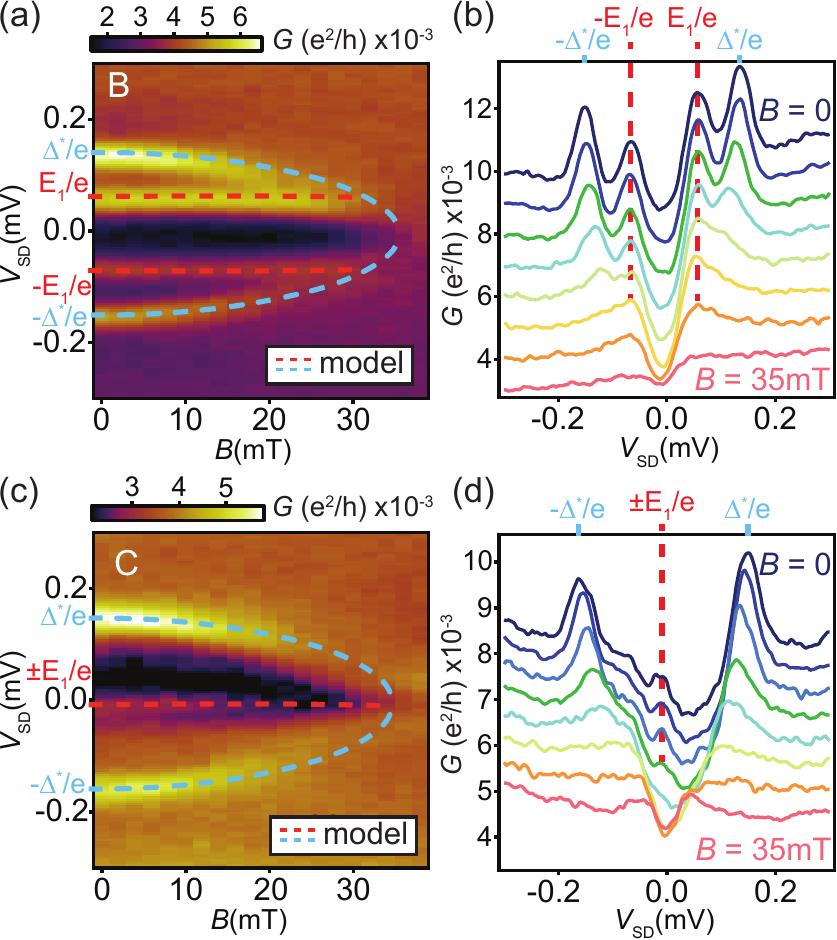}
\caption{\label{fig:SOI_ABS} $G$ as a function of $B$ (out of plane) and $V_{\text{SD}}$ for (a) region B ($V_{\rm{BG,B}} = \SI{-6.418}{\volt}$) and (c) region C ($V_{\rm{BG,C}} = \SI{-6.312}{\volt}$). The dashed blue and red lines are the superimposed modeled spectra. (b) and (d) show waterfall plots of the same data. In both regions we find a sub-gap resonance pinned to the constant energy $\pm$E$_{1}$, independent of $B$.}
\end{figure}
%
%
%
\begin{figure}[b]
\includegraphics[width=1\columnwidth]{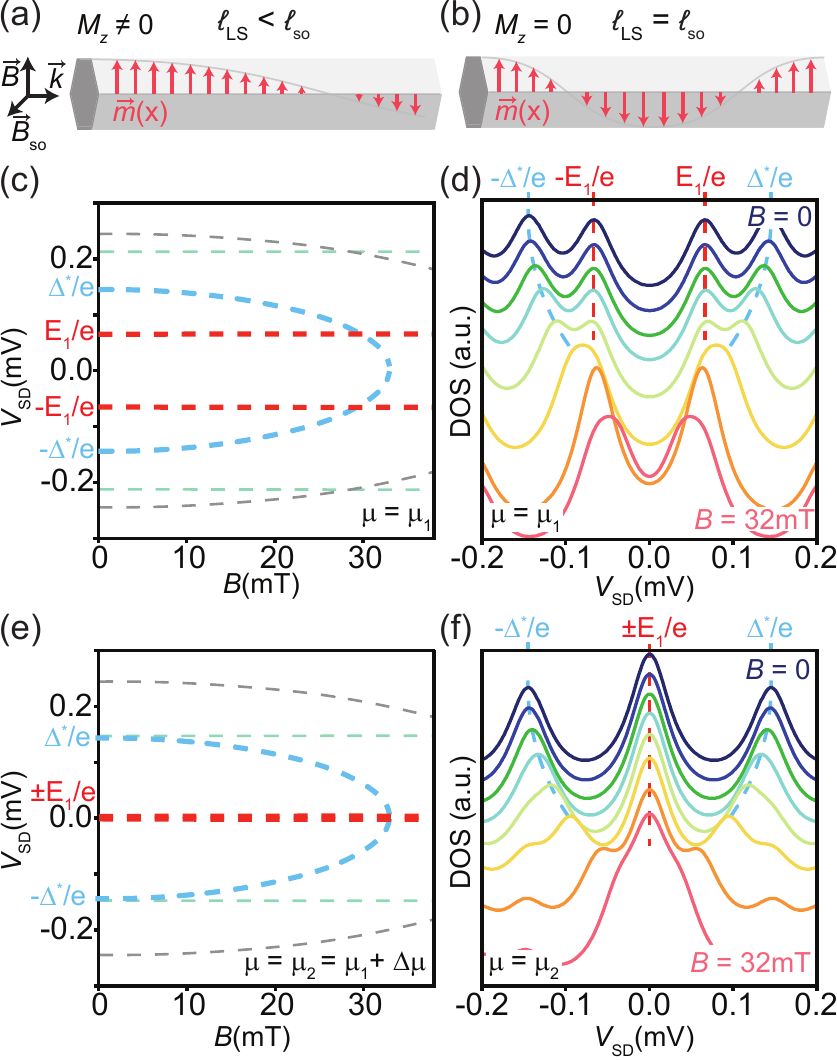}
\caption{\label{fig:theory} Schematic depiction of the spin texture for a given injected spin for (a) the non-commensurate case, $\ell_{\rm{LS}} < \ell_{\rm{so}}$, and (b) a commensurate case, $\ell_{\rm{LS}} = \ell_{\rm{so}} $, in a NW LS of length $\ell_{\rm{LS}}$. (c), (e) Calculated spectra as a function of $B$ and $V_{\text{SD}}$ for a commensurate case, reproducing the characteristics of  (c) region B at chemical potential $\mu_1$ and (e) for region C at $\mu_2 = \mu_1 + \Delta \mu$, respectively. In the calculations, the chemical potentials are chosen as $\mu_{SC}$~=~\SI{100}{meV}, $\mu_1 =$ \SI{10.8}{meV} and $\mu_2 =$ \SI{11.2}{meV}. In both cases a sub-gap state is pinned to a constant energy $\pm$E$_{1}$, independent of $B$. (d) and (f) show waterfall plots of the calculated DOS at the respective chemical potential using a constant resonance broadening of \SI{40}{\micro eV}.}
\end{figure}
%
In  Figs.~\ref{fig:SpectroABS}(c) - (f) we plot detailed measurements of the four regions A - D pointed out in Fig.~\ref{fig:SpectroABS}(b), including cross sections at the indicated $V_{BG}$. The characteristics of the respective sub-gap states are summarized  in Table~I in the Supplemental Material.
The sub-gap states detected in regions A and D show the standard characteristics of ABS or Yu-Shiba Rusinov states \cite{Eichler2007,Pillet2010,Lee2012,Chang2013,Jellinggaard2016,Kim2013,Gramich2017,Grove-Rasmussen2018}, expected for a strong coupling $\Gamma_S$ of the confined states to the SC ($\Gamma_S \geq \Delta^*$), possibly accompanied with a strong Kondo effect  \cite{Eichler2007}, as discussed in the Supplemental Material.\par
In stark contrast, the sub-gap resonances of regions B and C, shown in detail in Fig.~\ref{fig:SpectroABS}~(d) and (e), deviate from the standard picture, especially regarding the magnetic field dependence, discussed below in detail.\par
In region B we detect a second resonance at $E_1/e~\approx~\pm$~\SI{60}{\micro V}, which is gate tunable, but does not cross zero bias. Here, the gap edges remain visible at $\pm \Delta^*$. We note that we do not find a Kondo feature for this region in the normal state.\\
In region C, shown in Fig.~\ref{fig:SpectroABS}~(e), the resonance crosses zero energy, forming a small loop structure, best visible for negative $V_{\text{SD}}$. This loop structure is usually interpreted as the ground state transition from an ABS singlet to a spin 1/2 doublet \cite{Lee2013,Chang2013, Gramich2017}.
The gap at $\pm \Delta^*$ again remains visible, in addition, we find a resonance at $E_1/e \approx 0$, independent of $V_{\rm{BG}}$ on the scale of the corresponding charge state.\par
%
%
%
%
%
%
%
%
%
%
%
%
%
%
We now investigate the dependence of these sub-gap states on a magnetic field applied perpendicular to the substrate plane. The device remains in the superconducting state for $B < B_\text{C} \approx$ \SI{35}{\milli \tesla}, i.e. smaller than the critical magnetic field. In this field range, regions A and  D show a field dependence consistent with the standard picture of the sub-gap state energy decaying with the superconducting gap $\Delta^* (B)$, which vanishes at $B_\text{C}$, see  Fig.~S2~(a, c) in the Supplemental Material \cite{Chang2013,Heck2017}. Region D , in addition, shows a pronounced zero-bias peak, possibly related to the strong Kondo interactions.\par
%
%
%
%
%
%
In Figs.~\ref{fig:SOI_ABS}~(a) and (c), $G$ is plotted as a function of $B$ and $V_{\text{SD}}$ for regions B and C. For both regions we observe a closing of the superconducting gap $\Delta^*$, with increasing $B$. In addition, we find sub-gap resonances at constant energies $E_1 \approx \pm$~\SI{60}{\micro eV} (region B) and $E_1 \approx$~0 (region C), which are field-independent within the measurement accuracy, better seen in the waterfall plots in  Figs.~\ref{fig:SOI_ABS}~(b) and (d) respectively. Another resonance with similar characteristics is shown in Fig.~S3 in the Supplemental Material. In the vicinity of the critical field  ($B \approx B_C$), the states at constant energy merge with the conductance maxima at $\pm \Delta^*$.\par
The field independent sub-gap states can be understood intuitively by considering a classical magnetic moment $\vec{m}$ in a magnetic field $|\vec{B}_{\rm{so}}| = \frac{2\alpha p}{g^*\mu_{\rm{B}}}$ given by the Rashba SOI parameter $\alpha$ and the linear momentum $p$ at small external magnetic fields. Here, $g^*$ is the effective \emph{g}-factor and $\mu_{\rm{B}}$ the Bohr magneton. $\vec{m}$ precesses around $\vec{B}_{\rm{so}}$,
which translates into a characteristic length scale $\ell_{\rm{so}} = \frac{\pi \hbar}{\rm{m^*}\alpha}$ over which $\vec{m}$ rotates by a {\it full} period, using the reduced Planck constant $\hbar$ and the effective electron mass in ZB InAs, $\rm{m}^* \approx 0.023\rm{m}_e$.
At each reflection at the sharp boundaries of the finite length LS, the linear momentum is inverted, resulting in the inversion of $\vec{B}_{\rm{so}}$, so that $\vec{m}$ backtracks the precessional motion.  For a given injected orientation of $\vec{m}$, this leads to a static spin texture for each LS boundstate, with an effective magnetic moment $\vec{M} = \int_{\rm{0}}^{\rm{\ell_{\rm{LS}}}} \vec{m}\rm{(x)dx}$ depending on $\ell_{\rm{so}}$, and a shift in the resulting boundstate energy of $\Delta E = \vec{M} \cdot \vec{B}$ in a (weak) external magnetic field $\vec{B}$ defining the \emph{z}-direction.
For illustration we choose $\vec{B} \perp \vec{B}_{\rm{so}}$ and plot in Figures~\ref{fig:theory}~(a) and (b) the \emph{z}-component of $\vec{m}\rm{(x)}$ for $\ell_{\rm{LS}}$ being (a) non-commensurate and (b) commensurate to $\ell_\mathrm{{so}}$.
In the case of commensurability, i.e. $ \ell_{\rm{LS}} \approx  n\ell_{\mathrm{so}}$/2 $(n \in \mathbb{N})$, one finds $M_z \approx 0$, independent of the injected spin orientation, which results in field independent eigenenergies. We note that the components of $\vec{m}\rm{(x)}$ parallel to $\vec{B}_{\rm{so}}$ remain constant, so that the energy shifts due to fields $\vec{B} \parallel \vec{B}_{\rm{so}} $ are not affected. This intuitive picture can also be applied to superconducting boundstates and qualitatively accounts for the field independent sub-gap states observed here. A fully quantum mechanical description can be found in \cite{Reeg2018}, suggesting superposition states with equal weights of up and down spin projections in the commensurate case.
This interpretation holds as long as $B \ll B_{\rm{so}}$ and allows, for example, to extract a \textit{lower} bound for $E_{\rm{so}}$, since $\ell_{\rm{LS}} \geq \ell_{\rm{so}}$/2 ($n = 1$) for flat states, which yields $E_{\rm{so}} = \pi^2\hbar^2/2\rm{m}^* \ell^2_{\rm{so}}~\geq \pi^2\hbar^2/8\rm{m}^* \ell^2_{\rm{LS}} \geq$ ~\SI{85}{\micro e\volt}, corresponding to $\hbar\alpha \geq $~\SI{0.21}{eV\angstrom}. We note that this effective magnetic moment has strong implications for the definition and observed \emph{g}-factor values in confined electronic systems \cite{Flindt2006, Dmytruk2018}. 
\par
This qualitative picture is supported by a numerical tight binding model \cite{Reeg2018}.
We write the Hamiltonian of a one channel system in the Nambu basis as
\begin{equation}  \label{tight-binding}
\begin{aligned}
& H_{NW}= \sum_{x=1}^{\ell_{\rm{LS}}+\ell_{\rm{SC}}-1}\biggl[c^\dagger_x (2\hat{t}_{x}-\mu_x -\Delta_{Z,x}\sigma_z)c_x \\
& -\{c^\dagger_x (\hat{t}_{x} -i \alpha_{x} \sigma_{y}/2)c_{x+1}+\Delta_x c^\dagger_{x,\downarrow}c^\dagger_{x,\uparrow}+H.c.\}\biggr], 
\end{aligned}
\end{equation}
where $c_x=(c_{x,\uparrow},c_{x,\downarrow})^T$ and $c^\dagger_{x,\sigma}$ ($c_{x,\sigma}$) creates (annihilates) an electron with spin $\sigma$ at site $x$.
The boundary between both segments is at $x_{\rm{B}}=\ell_{\rm{LS}}-1/2$ where the Rashba coupling strength $\alpha_x=\widetilde{\alpha}\theta_{\rm{LS}}$, the Zeeman term $\Delta_{Z,x}=\Delta_Z\theta_{\rm{LS}}$, the hopping parameter $\hat{t}_{x}=t_{\rm{LS}}\theta_{\rm{LS}}+t_{\rm{SC}}\theta_{\rm{SC}}$, the chemical potential $\mu_x=\mu\theta_{\rm{LS}}+\mu_{SC}\theta_{\rm{SC}}$ and the pairing potential $\Delta_x=\Delta^*\sqrt{1-(\Delta_Z/\Delta_Z^c)^2}\theta_{\rm{SC}}$ change abruptly at the interface, with $\Delta_Z^c$ the critical magnetic field of the superconductor and the zero field pairing potential $\Delta^*$.
Here, we introduce the two heaviside functions $\theta_{\rm{LS}} = \theta(x_{\rm{B}}-x)$ and $\theta_{\rm{SC}} = \theta(x-x_{\rm{B}}) $, non-zero in the LS and below the SC, respectively. 
This configuration of parameters originates from previous studies suggesting that a strongly coupled SC drastically renormalizes the parameters of the NW \cite{Reeg2018,Reeg2018a}. This model is chosen here because it precludes the formation of a topological phase.
In order to model the experiment more realistically, we set up a Hamiltonian with three such parallel channels and a weak coupling in between. For the plots in Fig.~\ref{fig:theory} we set the model lattice constant to $a=$~\SI{5}{\nano \meter}, $\widetilde{\alpha} = \hbar\alpha/2a$ with $\hbar\alpha = $~\SI{0.26}{eV\angstrom}, $\ell_{\rm{LS}} = $~\SI{220}{\nano \meter} and the length of the SC $\ell_{\rm{SC}} = $~\SI{300}{\nano \meter} as in the experiment.
This results in the hopping strengths $t_{\rm{LS}} =$~\SI{100}{\milli e\volt} and $t_{\rm{SC}} = $~\SI{50}{\milli e\volt}. Furthermore, we set $\Delta^* =$~\SI{150}{\micro e\volt}, $\Delta_Z^c = $~\SI{35}{\milli \tesla} and the $g$-factor to $g = 15$, based on values extracted from the experiments.\par
In Fig.~\ref{fig:theory}~(c) and (e) we plot the eigenenergies obtained for the described model, and also use them as illustration in Fig.~\ref{fig:SOI_ABS}~(a) and (c). The model  reproduces all relevant characteristics of the experiment, namely that the resonances at the gap edges at $\pm \Delta^*$, corresponding to higher energy states, tend to zero for increasing $B$, whereas the sub-gap resonances at $\pm E_1$ do not depend on $B$. In the calculation the chemical potential within the LS as well as the SOI strength are adjusted to reproduce the characteristics of the measurement.
The absolute energy of the sub-gap states are determined by the chemical potential $\mu$ in the LS. By shifting $\mu$ by $\Delta \mu = \beta \Delta V_{\rm{BG}} \approx $~\SI{400}{\micro eV}, using the lever arm $\beta$ of the experiment, we can also account for states at $E_1 \approx 0$, see Fig.~\ref{fig:theory}~(e), which reproduces the measurements of region C in Fig.~\ref{fig:SOI_ABS}~(c). In the model, such flat states occur only for a sharp confinement, easily obtained at the integrated QD. At the SC interface, such a sharp confinement can occur due to a large step in the chemical potential in the case of a strong coupling between the SC and the NW.\par
In summary, we present tunnel spectroscopy measurements on discrete sub-gap resonances in a NW segment connected to one SC contact and using an integrated QD as a tunable tunnel barrier. We show that some of the sub-gap resonances can be qualitatively understood as standard sub-gap states, not explicitly taking into account SOI, except for an unknown effective \emph{g}-factor, with an excitation energy tending to zero with the closing of the gap for an increasing magnetic field. However, sub-gap states can also be pinned to a constant energy as a function of magnetic field, due to an equal superposition of both spin states in a NW LS of finite length. Such states can also reside at zero energy. We quantitatively reproduce the experiment with a theoretical model and use this effect to extract a lower bound of the SOI energy of ZB InAs NWs, namely $E_{\rm{so}}\geq$~\SI{85}{\micro e\volt}, consistent with previous results \cite{Fasth2007, Scheruebl2016}.\par
The authors thank C. Reeg, S. Hoffman, S. Diaz, M. Thakurathi, M. Nilsson, D. Indolese, G. Steffensen and J. Paaske for fruitful discussions. This research was supported by the Swiss National Science Foundation through a) a project grant entitled "Quantum Transport in Nanowires" granted to CS b) the National Center of Competence in Research Quantum Science and Technology (QSIT) and c) the QuantEra project SuperTop. We further acknowledge funding from the European Union's Horizon 2020 research and innovation program under grant agreement No 828948, project AndQC, the project QUSTEC as well as the Swiss Nanoscience Institute (SNI). D.C. acknowledges funding from the European Union’s Horizon 2020 research and innovation program (ERC Starting Grant, agreement No 757725). S.L., K.A.D. and C.T. acknowledge financial support by the Knut and Alice Wallenberg Foundation (KAW) and the Swedish Research Council (VR). All data in this publication are available in numerical form at \href{https://doi.org/10.5281/zenodo.3558308}{https://doi.org/10.5281/zenodo.3558308}

\bibliography{literature}
\newpage
\newpage
\onecolumngrid
%
%
\renewcommand{\thefigure}{S\arabic{figure}}
\setcounter{figure}{0}
%
%
\section*{\large{Supplemental Material:\\Magnetic field independent sub-gap states in hybrid Rashba nanowires}}
\section*{Additional data and analysis.}
\paragraph*{\bf{Kondo effect and visibility of superconducting gap.}} In Table \ref{tab:summary} we summarize the extracted characteristics of the resonances in regions A - D of Fig.~2 in the main text. For regions A and D we detect strong Kondo features in the normal state (see Fig.~2~(a), $B = \SI{50}{\milli \tesla}$) with a Kondo temperature of $T_\text{K} \approx$ \SIrange{200}{250}{\micro eV}, larger than the induced superconducting gap $\Delta^*$.
For region D we find a strong zero bias peak, probably related to Kondo physics in the superconducting state (see Fig.~2~(b), $B = 0$)  \cite{Eichler2007,Lee2012}.
We note that for A and D we do $\it{not}$ observe a conductance peak at the superconducting gap at $\pm \Delta^*$.\par
In contrast, we detect only a weak Kondo ridge for region C, with $T_\text{K} < \SI{25}{\micro eV}$ in the normal state and clearly pronounced peaks at the gap edges at $\pm \Delta^*$. For region B, we find no Kondo effect in the normal state and pronounced peaks at $\pm \Delta^*$.
\begin{table}[htb]
\begin{tabularx}{8.5cm}{|X|X|X|X|X|}
\hline
 Region: & A & D & C  & B \\
\hline
Kondo ridge in N-state & strong & strong & weak & none \\
\hline
$T_\text{K}$ & \SI{250}{\micro eV} & \SI{200}{\micro eV} & $<$\SI{25}{\micro eV} & NA \\
\hline
$\Delta^*$ edge visible & no & no & yes & yes \\
\hline

\end{tabularx}

\caption{Summary of characteristics of sub-gap states in regions A, D, C, B.\label{tab:summary}}

\end{table}

\begin{figure}[t!]
\includegraphics[width=0.45\columnwidth]{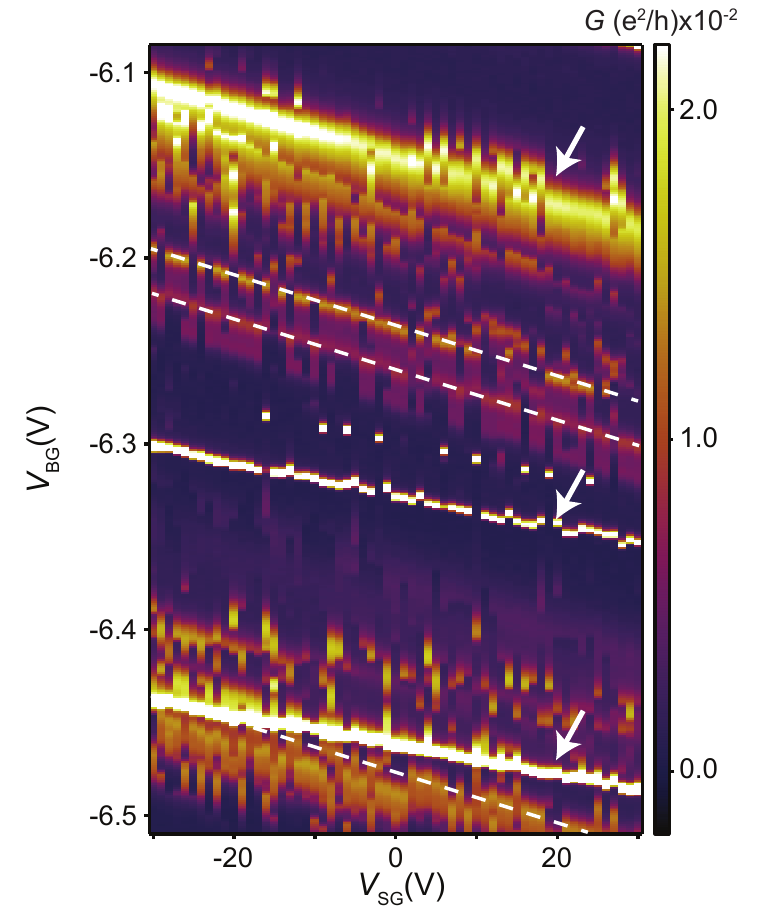}
\caption{\label{fig:sidegate} Differential conductance $G$ as a function of the sidegate voltage $V_{\mathrm{SG}}$ and the backgate voltage $V_{\mathrm{BG}}$ at $V_{\mathrm{SD}}~=~0$. Resonances of the integrated QD are labeled with white arrows. Dashed lines indicate lead resonances.}
\end{figure}
%
%
%
%
%
%
%
%
\paragraph*{\bf{Distinguishing QD and LS resonances.}} In  Fig.~\ref{fig:sidegate} we plot the differential conductance $G$ as a function of backgate voltage $V_{\mathrm{BG}}$ and sidegate voltage $V_{\mathrm{SG}}$. White arrows label the Coulomb blockade resonances of the integrated quantum dot, whereas the white dashed lines indicate the position of the LS resonances. These resonances can be distinguished by the different slope, i.e. different coupling capacitances to the two gates. In contrast, excited states of the integrated QD show the same slope as the fundamental resonance.\\\\
\paragraph*{\bf{Charge rearrangements in the data of Fig.~2}} In Fig.~2~(a, b) several charge rearrangements took place which we corrected by shifting the gate respectively. We note that no rearrangements occurred in the regimes discussed in the main text.\\\\
\begin{figure}[h!]
\includegraphics[width=0.48\columnwidth]{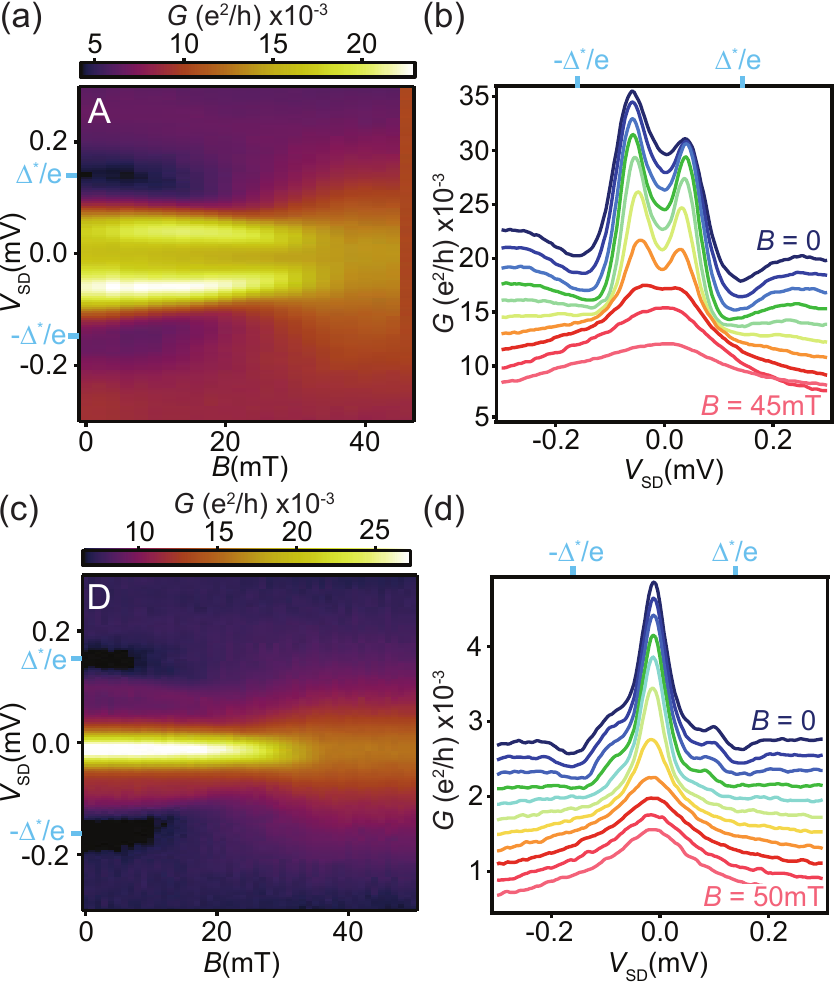}
\caption{\label{fig:Bfield_regionA_D} $G$ (color scale) as a function of $V_{\mathrm{SD}}$ and external magnetic field $B$ (out of plane) for (a) region A and, (c) D at the indicated $V_{\mathrm{BG}}$ (blue arrow) in Fig.~2~(c, f) respectively ($V_{\rm{BG,A}} = \SI{-6.471}{\volt}$, $V_{\rm{BG,D}} = \SI{-6.237}{\volt}$). (b) and (d) show a selection of the same data as a waterfall plot.}
\end{figure}
%
\paragraph*{\bf{Magnetic field dependence of the resonances in the regions A and D.}} In Fig.~\ref{fig:Bfield_regionA_D} we plot $G$ as a function of source drain voltage $V_{\mathrm{SD}}$ and external magnetic field $B$ for regions A and D respectively of Fig.~2 of the main text. In Fig.~\ref{fig:Bfield_regionA_D}~(a) we find the standard closing with increasing $B$ for the resonance in region A. For region D (see Fig.~\ref{fig:Bfield_regionA_D}~(c)) we find a pronounced zero bias peak which persists up to $B_{\mathrm{c}}$, probably related to a strong Kondo coupling.
\newpage
%
%
%
%
%
%
%
%
\paragraph*{\bf{Additional magnetic field independent sub-gap states in a different gate region.}}
In Fig.~\ref{fig:ABSflat3}~(a) we plot $G$ as a function of $V_{\text{SD}}$ and $B$ at a backgate voltage of $V_{\rm{BG}} = \SI{-6.11}{\volt}$, i.e. at a larger gate voltage than in the main text. At energies $\pm E_1/e$ we find again magnetic field independent sub-gap resonances. In addition, we observe peaks at the gap edges $\pm \Delta^*/e$ decreasing in energy as a function of $B$.  Fig.~\ref{fig:ABSflat3}~(b) shows a waterfall plot of the same measurement to highlight magnetic field independent sub-gap states at $\pm \Delta^*/e$ with red dashed lines.\par
%
%
%
\begin{figure}[h!]
\includegraphics[width=0.5\columnwidth]{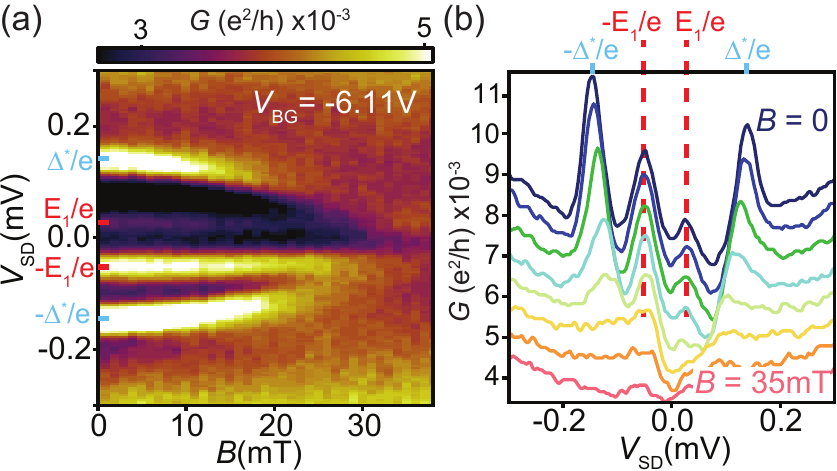}
\caption{\label{fig:ABSflat3} $G$ vs. $B$ (out of plane) and $V_{\mathrm{SD}}$ for a gate region not shown in the main text ($V_{\mathrm{BG}} = $ \SI{-6.11}{\volt}).  A sub-gap state stays pinned to constant energy $\pm$E$_{1}$, independent of $B$. (b) shows a waterfall plot of this region.}
\end{figure}
%
%
%
%
%
%
%
%
%
%
%
%
\section*{Additional numerical calculations for nanowire without spin-orbit interaction.}
\begin{figure}[h!]
\includegraphics[width=0.52\columnwidth]{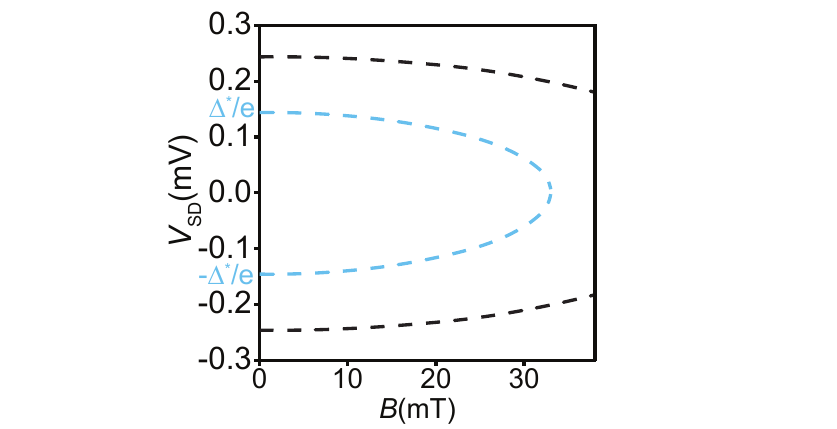}
\caption{\label{fig:no_SOI} Calculated eigenenergies of the NW system as a function of $B$ and $V_{\mathrm{SD}}$ with $\alpha = 0$ for the exact same parameters as used in Fig.~4~(c).}
\end{figure}

The physics discussed in the main text is based on the interplay between spin-orbit interaction (SOI) and the length of the lead segement, $\ell_{\mathrm{LS}}$. To show the direct relation Fig. \ref{fig:no_SOI} shows the eigenstates of the same system and parameter settings as in the Fig. 4 (c), except that we set the SOI strength $\alpha$ to zero. For these settings, there are no $B$ independent states, only spin degenerate bound states (due to the absence of SOI) that decrease in energy with increasing $B$ as does $\Delta^*$.


\section *{Bibliography of Supplemental Material}
[1] A. Eichler, M. Weiss, S. Oberholzer, C. Schönenberger, A. L. Yeyati, J. C. Cuevas, and A. Martin-Rodero, Physical Review Letters \textbf{99}, 126602 (2007).\par
[2] E. J. H. Lee, X. Jiang, R. Aguado, G. Katsaros, C. M. Lieber, and S. D. Franceschi, Physical Review Letters~\textbf{109}, 186802 (2012).

\end{document}